\def\a{\alpha}\def\b{\beta}\def\c{\chi}
\def\f{\phi}\def\h{\theta}
\def\k{\kappa}\def\l{\lambda}\def\m{\mu}\def\n{\nu}\def\q{\psi}\def\t{\tau}
\def\y{\eta}

\def\inf{\infty}\def\id{\equiv}\def\ha{{1\over 2}}
\def\app{\approx}

\def\mn{{\mu\nu}}

\def\coo{coordinates }

\def\pb{Poisson brackets }\def\db{Dirac brackets }

\def\des{de Sitter }

\def\tls{transformation laws }

\def\section#1{\bigskip\noindent{\bf#1}\smallskip}

\def\PL#1{Phys.\ Lett.\ {\bf#1}}

\def\PR#1{Phys.\ Rev.\ {\bf#1}}\def\CQG#1{Class.\ Quantum Grav.\ {\bf#1}}

\def\MPL#1{Mod.\ Phys.\ Lett.\ {\bf #1}}

\def\grq#1{{\tt gr-qc/#1}}\def\hep#1{{\tt hep-th/#1}}\def\arx#1{{\tt arXiv:#1}}

\def\ref#1{\medskip\everypar={\hangindent 2\parindent}#1}
\def\beginref{\begingroup
\bigskip
\centerline{\bf References}
\nobreak\noindent}
\def\endref{\par\endgroup}

\def\app{\approx}

%%%%%%%%%%%%%%%%%%%%%%%%%%%%%%%%%%%%%%%%%%%%%%%%%%%%%%%%%%%%%%%%
\magnification=1200\baselineskip=18pt

{\nopagenumbers
\line{\hfil July 2008}
\vskip60pt
\centerline{\bf Triply special relativity from six dimensions}
\vskip60pt
\centerline{
{\bf S. Mignemi}\footnote{$^\ddagger$}{e-mail:
smignemi@unica.it}}
\vskip10pt
\centerline {Dipartimento di Matematica, Universit\`a di Cagliari}
\centerline{viale Merello 92, 09123 Cagliari, Italy}
\centerline{and INFN, Sezione di Cagliari}
\vskip80pt
\centerline{\bf Abstract}

\vskip10pt
{\noindent
We show that the generalization of Doubly Special Relativity to a curved \des
background can be obtained starting from a six-dimensional spacetime
on which quadratic constraints on position and momentum \coo are imposed.
}
\vskip100pt\
P.A.C.S. Numbers: 11.30.Cp, 03.30.+p
\vfil\eject}

\section{1. Introduction}
Recently, the possibility of deriving Doubly Special Relativity (DSR) models from higher
dimensions has been discussed from several points of view [1-3].

As is well known, DSR models are based on the deformation of the standard energy-momentum
dispersion law of relativistic particles [4], that is assumed to be induced by quantum gravity
effects at scales near the Planck energy.
The deformation is realized by modifying the action of the Lorentz group
on momentum space through the introduction of a new observer-independent constant
$\k$, with the dimensions of energy (usually identified with the Planck energy).
In this framework, the transformation law of momenta becomes nonlinear, and that of positions
momentum dependent. Special relativity is recovered in the limit $\k\to\inf$.
Different choices of the deformed dispersion law correspond to different DSR models.
The various realizations can be obtained choosing different parametrizations
of a 5-dimensional momentum space, on which a quadratic constraint analogous to the
de Sitter hyperboloid of position space has been imposed [5].
This possibility has induced some authors to derive the DSR dynamics from a five-dimensional
model [1]. Another motivation for considering derivations from higher dimensions has been that
the \tls of some DSR models can be realized linearly in five dimensions [2].

In this paper, we wish to extend the derivation from higher dimensions of doubly special
relativity to the case of a de Sitter background spacetime (such a model has been called
triply special relativity (TSR) in [6]). The derivation is obtained starting from the phase space
of a six-dimensional particle, in a way analogous to that used in [7] to obtain Snyder spacetime
[8].

Generalizations of DSR to a (anti)-\des background have been discussed in different contexts in
[9-11,6], where various realizations of the deformed \des algebra have been proposed.
The main characteristic of these models is that both position and momentum \coo have
nonvanishing \pb between themselves. Nontrivial \pb between position coordinates arise naturally
in spacetime realizations of DSR models, and lead to noncommutative geometry at the quantum
level.
Moreover, nontrivial \pb between momenta (or better, translation generators), are a characteristic
of (anti)-\des spacetime.

As we shall see, the phase space of triply special relativity in four dimensions can be obtained
in the Hamiltonian formalism, by imposing on the position \coo a quadratic constraint
that generalizes the standard \des hyperboloid, and on the momentum \coo an analogous quadratic
constraint which yields the DSR structure on momentum space [5].
Enforcing both constraints on a 12-dimensional phase space, one is left with one first class and two
second class constraints. The hamiltonian reduction of the system gives rise to an 8-dimensional
phase space with a symplectic structure adapted to triply special relativity.
The specific model obtained depends on the choice of a gauge condition fixing the first class
constraint.

\section{2. The model}
We consider the action
$$S=\int d\t[\dot X_AP_A-(\l_1\f_1+\l_2\f_2+\l_3\f_3)],\eqno(1)$$
in a six-dimensional space of signature $(+,-,\dots,-)$,
with $\l_i$ Lagrange multipliers enforcing the constraints
$$\f_1=\ha(X_A^2+\a^2),\quad\f_2=\ha(P_A^2+\k^2),\quad\f_3=X_AP_A.\eqno(2)$$
Here $\a$ is the \des radius, while $\k$ is the energy scale of DSR.
The field equation read
$$\dot X_A=\l_2P_A+\l_3X_A,\qquad\dot P_A=-\l_1X_A-\l_3P_A.\eqno(3)$$
The constraints $\f_1\app\f_2\app0$ enforce the de Sitter structure of position and momentum [5]
space, while $\f_3\app0$ can be considered as a secondary constraint following from the other two.

The constraints satisfy the algebra
$$\{\f_1,\f_2\}\app0,\quad\{\f_1,\f_3\}\app-\a^2,\quad\{\f_2,\f_3\}\app\k^2.\eqno(4)$$
It is easy to see that they split into one first class constraint $\q$ and two second class
constraints $\c_i$. The original 12-dimensional phase space can therefore be reduced to 8 independent
coordinates.
By taking linear combination of the original constraints, one can define
$$\q=\ha(\k^2X_A^2+\a^2P_A^2+2\a^2\k^2),\eqno(5)$$
$$\c_1=X_AP_A,\qquad\c_2=\ha(\k^2X_A^2-\a^2P_A^2),\eqno(6)$$
with
$$C_{12}\id\{\c_1,\c_2\}\app-2\a^2\k^2.\eqno(7)$$
In terms of the new variables, the action reads
$$S=\int d\t\left[\dot X_AP_A-\ha\left({\l_1\over\k^2}+{\l_2\over\a^2}\right)\q
-\ha\left({\l_1\over\k^2}-{\l_2\over\a^2}\right)\c_1-\l_3\c_3\right].\eqno(8)$$
The compatibility conditions imply that the coefficient of the secondary constraint vanish, and hence
$$\a^2\l_1=\k^2\l_2,\qquad\l_3=0.\eqno(9)$$

Defining the Dirac brackets of two functions in phase space
as $\{A,B\}^*=\{A,B\}-\{A,\c_\a\}C^{\a\b}\{\c_\b,B\}$,
with $C^{\a\b}$ the inverse of $C_{\a\b}$, one has
$$\eqalignno{\{X_A,X_B\}^*&\approx{1\over2\k^2}(X_AP_B-X_BP_A),&\cr
\{P_A,P_B\}^*&\approx{1\over2\a^2}(X_AP_B-X_BP_A),&\cr
\{X_A,P_B\}^*&\approx\y_{AB}+{1\over2\a^2}X_AX_B+{1\over2\k^2}P_AP_B.&(10)}$$
The Dirac brackets have a structure analogous to that of the known DSR model in \des space [9-11,6].
In particular the \pb of positions and of momenta are proportional to the classical generators
of Lorentz transformations, while the position-momentum brackets contain terms quadratic
in positions and momenta.

However, in order to complete the Hamiltonian reduction and obtain a 8-dimensional phase space
adapted to a four-dimensional particle, one must still fix a gauge condition for the first-class
constraint.
According to [1], this choice singles out a specific model between the possible realizations of DSR
in \des space.

We give two examples of gauge conditions: The simplest is $\h\id P_5=0$.
It follows
$$P_4=\sqrt{P_\m^2+\k^2}\ ,\qquad X_4={X_\m P_\m\over\sqrt{P_\m^2+\k^2}}\ ,
\qquad X_5=\sqrt{{(X_\m^2+\a^2)(P_\m^2+\k^2)-(X_\m P_\m)^2\over P_\m^2+\k^2}}\ ,$$
where the greek indices run from 0 to 3.
All the constraint are now second class and the matrix $C_{\a\b}$ reads
$$C_{\a\b}\approx\pmatrix{0&0&0&-\k^2X_5\cr0&0&-2\a^2\k^2&-\k^2X_5\cr0&2\a^2\k^2&0&0\cr
\k^2X_5&\k^2X_5&0&0\cr},\eqno(11)$$
with inverse
$$C^{\a\b}\approx\pmatrix{0&0&-{1\over2\a^2\k^2}&{1\over\k^2X_5}\cr0&0&{1\over2\a^2\k^2}
&0\cr{1\over2\a^2\k^2}&-{1\over2\a^2\k^2}&0&0\cr-{1\over\k^2X_5}&0&0&0\cr}.\eqno(12)$$
The Dirac brackets for the 4-dimensional coordinates in this gauge are given by
$$\eqalignno{\{X_\m,X_\n\}^*&\approx{1\over\k^2}(X_\m P_\n-X_\n P_\m),&\cr
\{P_\m,P_\n\}^*&\approx0&\cr
\{X_\m,P_\n\}^*&\approx\y_\mn+{1\over\k^2}P_\m P_\n.&(13)}$$
These are the \pb associated with the Snyder model [8], which were obtained in [7] in a
slightly different way. Likewise, the gauge $X_5=0$, would have produced the \des algebra.

Another possible choice of gauge is $\h\id{X_5\over\a}+{P_5\over\k}=0$.
In this case, the matrix $C_{\a\b}$ reads
$$C_{\a\b}\approx\pmatrix{0&0&0&2\k X_5\cr0&0&-2\a^2\k^2&0\cr0&2\a^2\k^2&0&-{2X_5\over\a}
\cr-2\k X_5&0&{2X_5\over\a}&0\cr},\eqno(14)$$
with inverse
$$C^{\a\b}\approx\pmatrix{0&-{1\over2\a^3\k^3}&0&-{1\over2\k X_5}\cr{1\over2\a^3\k^3}&0
&{1\over2\a^2\k^2}&0\cr0&-{1\over2\a^2\k^2}&0&0\cr{1\over2\k X_5}&0&0&0\cr},\eqno(15)$$
and the Dirac brackets for the 4-dimensional coordinates are
$$\eqalignno{\{X_\m,X_\n\}^*&\approx{1\over2\k^2}(X_\m P_\n-X_\n P_\m),&\cr
\{P_\m,P_\n\}^*&\approx{1\over2\a^2}(X_\m P_\n-X_\m P_\n),&\cr
\{X_\m,P_\n\}^*&\approx\y_\mn+{1\over2\a^2}X_\m X_\n+{1\over2\k^2}P_\m P_\n+
{1\over\a\k}P_\m X_\n.&(16)}$$
These are the commutation relations proposed in [6] for TSR, in a representation in which
the Lorentz generators maintain the classical form $J_\mn=X_\m P_\n-X_\n P_\m$.

The previous gauge conditions can be generalized to
$\h\id\m_1{X_5\over\a}+\m_2{P_5\over\k}=0$, with generic coefficients $\m_1$, $\m_2$,
normalized so that $\m_1^2+\m_2^2=1$.
Going through the same steps as before, one obtains the \db
$$\eqalignno{\{X_\m,X_\n\}^*&\approx{\cos^2\m\over\k^2}\ (X_\m P_\n-X_\n P_\m),&\cr
\{P_\m,P_\n\}^*&\approx{\sin^2\m\over\a^2}\ (X_\m P_\n-X_\n P_\m),&\cr
\{X_\m,P_\n\}^*&\approx\y_\mn+{\sin^2\m\over\a^2}\ X_\m X_\n+{\cos^2\m\over\k^2}\
P_\m P_\n+{2\sin\m\cos\m\over\a\k}\ P_\m X_\n,&(17)}$$
where we have defined $\m_1=\cos\m$, $\m_2=\sin\m$.

Once one has performed the reduction of the phase space, the 4-dimensional dynamics can
be studied in a standard way by imposing a Hamiltonian
constraint $P_\m^2=M^2$, or equivalently $P_4^2+P_5^2=M^2+\k^2$ [1].

According to [1], it should also be possible to define a set of 4-dimensional \coo commuting
with all the constraints, in terms of the 6-dimensional ones. These \coo would satisfy \pb
identical to the \db obtained above. Unfortunately, we were not able to find an explicit
expression for them.

\section{3. Conclusion}

A six-dimensional constrained system has been reduced to four dimensions, giving rise to
Dirac brackets compatible with DSR in a \des background (TSR). In limit cases one obtains the
Snyder model or \des special relativity. The specific four-dimensional model obtained depends
on a choice of gauge (i.e.\ of  projection from six dimensions). In particular, we have been
able to recover the phase space of the model proposed in ref.\ [6]. Of course, more complicated
TSR models [9-11] can be obtained with less simple gauge choices.
Clearly, the various models are equivalent from a six dimensional point of view, but not in
a four-dimensional perspective.

\beginref
\ref [1] F. Girelli, T. Konopka, J. Kowalski-Glikman, E.R. Livine, \PR{D73}, 045009 (2006).
\ref [2] A.A. Deriglazov, B.F. Rizzuti, \PR{D71}, 123515 (2005).
\ref [3] S. Mignemi, \arx{0711.4053}.
\ref [4] For a review, see J. Kowalski-Glikman, \hep{0405273};
G. Amelino-Camelia, \grq{0412136}.
\ref [5] J. Kowalski-Glikman, \PL{B547}, 291 (2002);
J. Kowalski-Glikman, S. Nowak, \CQG{20}, 4799 (2003).
\ref [6] J. Kowalski-Glikman and L. Smolin, \PR{D70}, 065020 (2004).
\ref [7] J.M. Romero and A. Zamora, \PR{D70}, 105006 (2004).
\ref [8] H.S. Snyder, \PR{71}, 38 (1947).
\ref [9] S. Mignemi, \MPL{A18}, 643 (2003).
\ref [10] A. Ballesteros, F.J. Herranz, N.R. Bruno, \hep{0409295}.
\ref [11] S. Mignemi, \arx{0802.1129}.

\endref
\end